\documentclass[conference]{IEEEtran}
\IEEEoverridecommandlockouts
\usepackage{cite}
\usepackage{amsmath,amssymb,amsfonts}

\usepackage{graphicx}
\usepackage{textcomp}
\usepackage{xcolor}

\usepackage{algorithm}
\usepackage{algpseudocode}
\usepackage{float}

\usepackage{tikz}
\usetikzlibrary{arrows.meta,positioning,shapes.geometric,fit,calc}
\definecolor{softblue}{RGB}{222,235,247}
\definecolor{softgreen}{RGB}{226,239,218}
\definecolor{softorange}{RGB}{254,230,206}
\definecolor{softgray}{RGB}{245,245,245}

\usepackage{tikz}
\usetikzlibrary{arrows.meta,positioning,fit}

\usepackage{url} % good line breaks for long URLs
\usepackage[
  pdftex,
  colorlinks=true,    % set to false or use hidelinks for camera-ready
  linkcolor=black,    % internal links (sections, figs)
  citecolor=blue,     % citation numbers [1], [2], ...
  urlcolor=blue       % URLs/DOIs
]{hyperref}

\def\BibTeX{{\rm B\kern-.05em{\sc i\kern-.025em b}\kern-.08em
    T\kern-.1667em\lower.7ex\hbox{E}\kern-.125emX}}
\begin{document}

\title{Optimizing Anonymity and Efficiency: A Critical Review of Path Selection Strategies in Tor
\thanks{}}

% Replace your current \author{...} with this compact block
\author{
\IEEEauthorblockN{Siddique Abubakr Muntaka \quad Jacques Bou Abdo}
\IEEEauthorblockA{School of Information Technology, University of Cincinnati, Cincinnati, OH, USA \\
\{muntaksr@mail.uc.edu, bouabdjs@ucmail.uc.edu\}}
}

\maketitle

%\iffalse
% === IEEE accepted-manuscript notice (ARXIV VERSION, BOXED) ===
\noindent\begingroup
\setlength{\fboxsep}{8pt}% padding inside the box
\color{red}% frame color
\fbox{%
  \begin{minipage}{0.97\linewidth}\footnotesize\color{black}
  \textbf{IEEE Copyright Notice—}\\[0.2em]
  © 2025 IEEE. Personal use of this material is permitted. Permission from IEEE must be obtained
  for all other uses, in any current or future media, including reprinting/republishing this material
  for advertising or promotional purposes, creating new collective works, for resale or redistribution
  to servers or lists, or reuse of any copyrighted component of this work in other works.\\[0.3em]
  \textit{To appear in:} Proc.\ ACS/IEEE AICCSA 2025, Doha, Qatar, Oct 19–22, 2025.\quad
  \textit{DOI:} (to be assigned)
  \end{minipage}%
}%
\endgroup\par\medskip
%\fi

\begin{abstract}
The Onion Router (Tor) relies on path selection algorithms to balance performance and anonymity by determining how traffic flows through its relay network. As Tor scales and usage patterns evolve, default strategies such as bandwidth-weighted random selection and persistent guard nodes face increasing performance limitations. This study presents a comparative evaluation of five path selection strategies: Random, Guard, Congestion-Aware, and two Geographic approaches (Diversity Driven and Latency-Optimized), herein referred to as Geo-Latency and Geo-Diversity, using a high-fidelity simulation model inspired by TorPS (Tor Path Simulator). Experiments were conducted across five network scales, simulating 37,500 circuits under realistic relay conditions. Results show that Geographic (Latency-Optimized) consistently achieved the lowest latency (40.0 ms) and highest efficiency, while Congestion-Aware strategies delivered the best throughput, outperforming the baseline by up to 42\%. Guard nodes maintained stable routing but exhibited latency increases under larger networks. No single method proved optimal across all scenarios, but each revealed clear strengths for specific use cases. These findings demonstrate that targeted path selection can significantly improve Tor’s performance without compromising anonymity, providing guidance for optimizing circuit construction in future development and deployments.
\end{abstract}


\section{Introduction}
% --- Background and Context ---
The Tor network has become a critical tool for online privacy \cite{jain2025unmasking}, just like other anonymity networks like Invisible Internet Project (I2P) \cite{abdo2023modeling}, serving millions of users worldwide through its unique onion routing architecture \cite{muntaka2025mapping} \cite{muntaka2025resilience}. At the heart of this system lies the path selection mechanism, which determines how traffic is routed through volunteer-operated relays while maintaining user anonymity \cite{shoker2021tormass}. Current implementations primarily employ bandwidth-weighted random selection with persistent guard nodes, as reported in \cite{imani2018guard}, an approach designed to balance security and load distribution. However, these methods were developed when Tor's primary use case was anonymous web browsing \cite{aminuddin2023rise}, creating potential mismatches with modern usage patterns, including real-time communication and high-bandwidth applications.

% --- Problem Statement and Gap ---
In recent years, there has been a growing recognition of performance limitations in Tor's routing mechanisms. Users frequently experience significant latency and throughput constraints, which research suggests may discourage adopting privacy-preserving technologies \cite{panchenko2012improving}. This tension between performance and anonymity represents a fundamental challenge in anonymous communication systems like Tor. While the security properties of Tor's design have been extensively studied, investigation of path selection optimization remains comparatively limited in the literature.

% --- Existing Work and Open Challenges ---
Several promising alternative approaches have emerged in the literature, including congestion-aware routing \cite{wang2012congestion} and geographic optimization strategies. These proposals address specific performance bottlenecks while maintaining Tor's security guarantees. However, existing studies often focus on individual algorithms in isolation, leaving open questions about their relative effectiveness and trade-offs. Furthermore, there is a limited understanding of how these methods scale with network size and varying user loads, which are critical considerations for a rapidly growing anonymity network. Recognizing this gap, our work's key contributions include:

\begin{itemize}
    \item \textbf{Comprehensive Comparative Evaluation:} A comparison of five path selection strategies: Random, Guard, Congestion-Aware, and two Geographic approaches across five network scales under realistic conditions.

    \item \textbf{High-Fidelity Simulation Model:} A simulation framework that incorporates realistic relay attributes, including geographic distribution and dynamic congestion, providing a robust evaluation environment beyond simplified models.

    \item \textbf{Targeted Performance Insights:} The study reveals the clear strengths and trade-offs of each strategy, identifying the optimal choice for latency-critical and throughput-oriented applications.

    \item \textbf{Bridging Theory and Practice:} Practical, data-driven insights for network operators and developers to inform decisions on optimizing circuit construction.
\end{itemize}

The remainder of this paper is organized as follows: Section~II reviews related work. Section~III presents our methodology, including the simulation framework, network model, and path selection strategies. Section~IV reports our results and analysis. Sections~V and VI conclude the paper with a discussion of findings, limitations, and future research directions.

\section{Related Work}
\begin{figure}[htbp]
\centering
\includegraphics[width=0.40\textwidth]{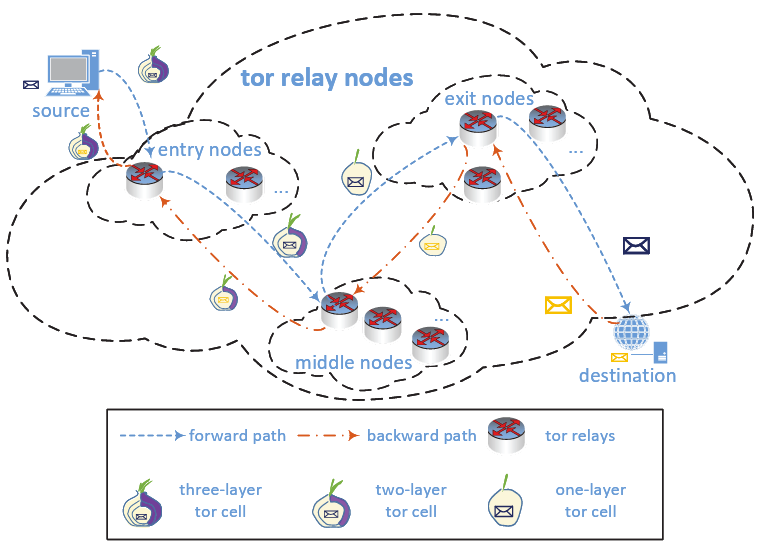}
\caption{The Tor circuit is built upon three Tor relays (entry, middle, exit) \cite{zhang2022less}.}
\label{fig:torpathselectiondiagram}
\end{figure}

Tor’s path selection mechanism constructs circuits through three types of relays; entry (guard), middle, and exit nodes, which collectively form the foundation of its anonymity guarantees. As shown in Figure~\ref{fig:torpathselectiondiagram}, data is routed through a multi-hop encrypted circuit where each relay decrypts a single encryption layer to obscure source and destination identities \cite{muntaka2025resilience}. While this structure preserves anonymity, it is vulnerable to traffic correlation attacks that can compromise user privacy by observing both ends of a circuit. Moreover, the choice of path selection strategy directly influences performance, creating a trade-off between anonymity and efficiency.

The fundamental tension between anonymity preservation and performance optimization has driven the evolution of path selection research in Tor. Early work, including that by Syverson et al.~\cite{syverson2001towards} and Wang et al.~\cite{wang2012congestion}, introduced the bandwidth-weighted random selection technique, which aims to distribute load across the network while maintaining routing unpredictability. However, this approach's performance limitations, especially under real-world conditions, have encouraged researchers to explore more adaptive and context-aware alternatives. Guard node selection was proposed to mitigate risks from adversarial entry points. Loesing et al.~\cite{loesing2010case} demonstrated how persistent guard nodes reduce vulnerability to certain attacks by minimizing the frequency of entry-node changes. Imani et al.~\cite{imani2018guard} extended this by highlighting improvements in performance stability. However, Wan et al.~\cite{wan2019guard} revealed critical weaknesses in guard placement, demonstrating that adversaries can strategically position malicious guards. Their findings emphasized the need to balance guard stability with security-aware placement.

To address performance bottlenecks, congestion-aware routing techniques have emerged as promising alternatives. Wang et al.~\cite{wang2012congestion} introduced a congestion-aware routing (CAR) framework that dynamically avoids overloaded relays by monitoring real-time network conditions. Building on this, Geddes et al. proposed avoiding bottleneck relay algorithm (ABRA), a congestion-aware relay selection algorithm that improved median throughput by almost 20\% in Tor by dynamically avoiding bottleneck relays, surpassing the default Tor relay selection method \cite{geddes2016abra}. This demonstrates the benefits of integrating real-time congestion feedback into circuit construction.

Geographic optimization strategies offer another approach to improving performance. Milajerdi and Kharrazi~\cite{milajerdi2015composite} introduced a composite-metric selection model that factors in both geographic distance and latency. Their results showed meaningful reductions in end-to-end delays while maintaining anonymity. Similarly, Zhang et al. analyzed vulnerabilities in autonomous system aware (AS-aware) selection methods and proposed distance-aware alternatives that mitigate correlation attacks ~\cite{zhang2022less}. They identified that reliance on route inference and content delivery networks (CDN) induced distortions that weaken traditional AS-based protections.

The growing complexity of Tor’s performance landscape has led to the development of adaptive congestion control mechanisms. For instance, PredicTor, proposed by Fiedler et al.~\cite{fiedler2020predictor}, applies distributed model predictive control to dynamically schedule traffic across relays. In contrast, Barton et al. took a global and location-agnostic approach to network analysis. Their study proposed using a machine learning classifier to predict path performance before constructing a Tor circuit ~\cite{barton2025predictor}. By exchanging predicted queue states between adjacent nodes, system achieved improved latency, fairness, and congestion avoidance while preserving Tor's anonymity properties. Barton et al.~\cite{barton2025predictor} also explored predictive algorithms to anticipate network behavior, while Imani et al.~\cite{imani2019modified} proposed relay selection methods that utilized historical performance data to enhance circuit quality. Although these intelligent techniques showed improve performance metrics, their possible lack of transparency raises concerns regarding trust and auditability in anonymity networks.

Hybrid path selection strategies have also been explored. Kiran et al.~\cite{kiran2017client} developed client requirement-based selection mechanisms that adjust dynamically according to application needs and network state. Their work supports the viability of context-aware path selection that optimizes for task-specific demands while preserving anonymity guarantees.

Despite these advances, several challenges remain. Most studies evaluate single algorithms in isolation and lack comparative assessments across diverse operational scenarios. Additionally, the combined effects of multiple selection strategies on network-wide behavior are underexplored. There is a clear need for holistic evaluation frameworks that can account for variation in user demands, network scale, and relay characteristics.

This study addresses these gaps through the design of a high-fidelity simulation that compares five path selection algorithms under realistic network conditions. Unlike prior models that rely on static or simplified assumptions, our framework incorporates relay behavior, geographic distribution, and dynamic congestion. We extend the evaluation beyond basic performance metrics by assessing scalability, efficiency, and multi-dimensional trade-offs, thereby offering a more comprehensive understanding of Tor’s path selection design space.

\section{Methodology}

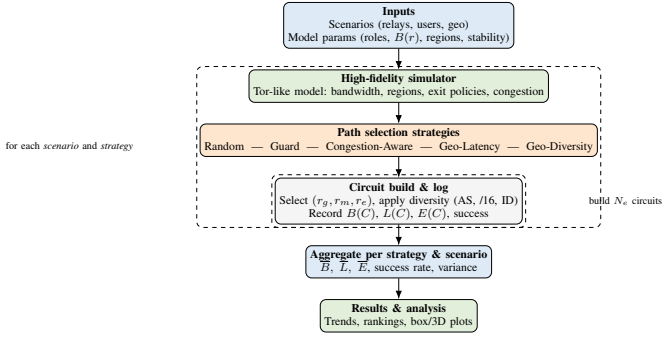
\begin{figure}[t]
\centering
\resizebox{\columnwidth}{!}{%
\begin{tikzpicture}[
  font=\footnotesize,
  node distance=5mm and 8mm,
  >=Latex,
  block/.style={rectangle, rounded corners, draw, semithick, align=center, inner sep=3pt, minimum height=7mm},
  b1/.style={block, fill=softblue},
  b2/.style={block, fill=softgreen},
  b3/.style={block, fill=softorange},
  b4/.style={block, fill=softgray},
  loopbox/.style={draw, dashed, rounded corners, inner sep=2pt}
]

% Nodes (top to bottom pipeline)
\node[b1] (inputs) {\textbf{Inputs}\\ Scenarios (relays, users, geo)\\ Model params (roles, $B(r)$, regions, stability)};
\node[b2, below=of inputs] (sim) {\textbf{High-fidelity simulator}\\ Tor-like model: bandwidth, regions, exit policies, congestion};
\node[b3, below=of sim] (strat) {\textbf{Path selection strategies}\\ Random \;|\; Guard \;|\; Congestion-Aware \;|\; Geo-Latency \;|\; Geo-Diversity};
\node[b4, below=of strat] (circuit) {\textbf{Circuit build \& log}\\ Select $(r_g,r_m,r_e)$, apply diversity (AS, /16, ID)\\ Record $B(C)$, $L(C)$, $E(C)$, success};
\node[b1, below=of circuit] (agg) {\textbf{Aggregate per strategy \& scenario}\\ $\overline{B},\ \overline{L},\ \overline{E}$, success rate, variance};
\node[b2, below=of agg] (out) {\textbf{Results \& analysis}\\ Trends, rankings, box/3D plots};

% Arrows
\draw[->,semithick] (inputs) -- (sim);
\draw[->,semithick] (sim) -- (strat);
\draw[->,semithick] (strat) -- (circuit);
\draw[->,semithick] (circuit) -- (agg);
\draw[->,semithick] (agg) -- (out);

% Minimal loop annotations (no clutter)
\node[loopbox, fit=(sim)(strat)(circuit), label={[xshift=-1.4cm]left:\scriptsize for each \textit{scenario} and \textit{strategy}}] (L1) {};
\node[loopbox, fit=(circuit), label={[xshift=1.4cm]right:\scriptsize build $N_e$ circuits}] (L2) {};

\end{tikzpicture}%
}
\caption{General methodology block diagram of the proposed evaluation pipeline.}
\label{fig:methodology_block}
\end{figure}

As shown in Figure ~ \ref{fig:methodology_block}, We implement a high-fidelity Tor path simulator in Python inspired by open-source TorPS (Tor Path Selector Simulator \cite{johnson2013users}). Our simulator models Tor's relay (nodes) network with attributes; bandwidth capacity, region-based latency, node stability, dynamic congestion, and exit policies, and evaluates five path selection strategies across five scaling scenarios.

% Addign Block Diagram

% --- Simple, colorful, compact methodology block diagram (TikZ) ---

\subsection{Network Model and Circuit Definition}
We generate network topologies containing $N$ relays by assigning each node a role as a guard, middle, and exit node in proportions (15\% guards, 15\% exits, 70\% middles) following Johnson et al \cite{johnson2013users} approach of using consensus documents from the live Tor network. This distribution also reflects a simplified but representative snapshot of the Tor network’s relay role proportions, acknowledging that actual distributions fluctuate over time due to node churn (joining and leaving) and volunteer dynamics as supported in the scholarly work of \cite{rahalkar2022analyzing}. For each relay $r$, we sample its bandwidth $B(r)$ (in KB/s) from a role-specific log-normal distribution, assign it to one of four geographic regions (North America, Europe, Asia, Rest of World) according to the probabilities (30\%, 40\%, 20\%, 10\% respectively), and determine its uptime and stability via exponential distributions with role-specific parameters. Each relay maintains a congestion metric $c(r)\in[0,1]$ updated periodically to reflect simulated load, and exit policies permitting traffic on common ports (80, 443) for exit-flagged relays.

A Tor circuit is defined as an ordered triple $C=(r_g,r_m,r_e)$ of guard, middle, and exit relays. Its effective throughput is given by the bottleneck bandwidth,
\begin{equation*}
 B(C) = \min\{B(r_g),B(r_m),B(r_e)\},
\end{equation*}
and its end-to-end latency by the sum of inter-hop delays,
\begin{equation*}
 L(C) = d(r_g, r_m) + d(r_m, r_e),
\end{equation*}
where $d(\cdot,\cdot)$ maps region pairs to representative latencies. Circuit efficiency is computed as $E(C) = B(C)/(L(C)+1)$ to evaluate overall performance.

\subsection{Path Selection Strategies}
We evaluate the five algorithms, as found in Algorithm \ref{alg:tor-evaluation} (Tor Path Selection Strategy Evaluation). \textit{Random (Bandwidth-Weighted)} selects each hop by sampling eligible relays with probability proportional to their bandwidth, $P(i)=B(i)/\sum_jB(j)$. \textit{Guard (Persistent Guards)} establishes a persistent list of three entry guards chosen by composite stability and bandwidth scores, rotating by least used first, with subsequent hops using bandwidth-weighted random sampling. \textit{Congestion-Aware} excludes relays with $c(r)\ge0.70$ and scores remainder by $B(r)\times[1-c(r)]$, selecting from top-quartile candidates. \textit{Geographic (Latency-Optimized)} identifies region triples $(R_g,R_m,R_e)$ minimizing total inter-region latencies, then applies bandwidth weighting within selected regions. \textit{Geographic (Diversity-Optimized)} maximizes geographic separation by preferring distinct regions while maintaining bandwidth weighting for intra-region selection.

%Add Algorithm
\begin{algorithm}[t]
\caption{Tor Path Selection Strategy Evaluation}
\label{alg:tor-evaluation}
\begin{algorithmic}[1]
\algrenewcommand\alglinenumber[1]{\footnotesize #1}
\Require Set of strategies $S = \{$Random, Guard, Congestion-Aware, Geo-Latency, Geo-Diversity$\}$
\Require Experimental scenarios $E$ with network and traffic parameters
\Ensure Performance metrics per strategy per scenario
\State Initialize simulation environment with Tor model
\ForAll{scenario $e \in E$}
    \State Configure relay count, user load, geo-distribution
    \State Generate network topology with relay attributes
    \ForAll{strategy $s \in S$}
        \State Initialize strategy-specific parameters
        \For{$i = 1$ to $N_e$} \Comment{$N_e$ circuits per scenario}
            \State Select path $(r_g, r_m, r_e)$ using $s$
            \State Apply diversity constraints (AS, subnet, ID)
            \State Compute $B(C)$, $L(C)$, $E(C)$, build time
            \State Log success/failure and performance data
        \EndFor
        \State Aggregate: $\overline{B}$, $\overline{L}$, $\overline{E}$, success rate, variance
        \State Trigger garbage collection
    \EndFor
\EndFor
\State Export results to JSON
\end{algorithmic}
\end{algorithm}

\subsection{Experimental Design}
The simulation constructs network topologies in batches of 500-2500 relays for memory efficiency. For each scenario-strategy combination, we generate circuits representing user traffic samples: 2,500 circuits for Scenario 1, 5,000 for Scenario 2, and 10,000 for Scenarios 3 to 5. Each circuit creation records throughput $B(C)$, latency $L(C)$, efficiency $E(C)$, success status, and build time. Diversity constraints enforce unique node IDs, autonomous systems, and /16 subnets per circuit path. Memory management includes periodic garbage collection between algorithms.

\subsection{Experimental Scenarios and Metrics}
We evaluate five experimental scenarios designed to vary client load and network scale systematically: (1) 250,000 users on 10,000 relays, (2) 500,000 users on 10,000 relays, (3) 1,000,000 users on 10,000 relays, (4) 1,000,000 users on 20,000 relays, and (5) 1,000,000 users on 50,000 relays. Each configuration enables targeted observation of scalability effects and stress thresholds in the Tor-like network model. The primary evaluation metrics include mean throughput $\overline{B}$, mean latency $\overline{L}$, efficiency score $\overline{E}$, circuit success rate, and performance consistency measured via standard deviation. In addition, secondary analyses focus on emergent scalability trends, specialization of path selection strategies under varying conditions, and the interplay between performance gains and anonymity-preserving constraints.

\section{Results}\label{sec4}

This section presents a comprehensive analysis of the five Tor path selection algorithms: Random, Guard, Congestion-Aware, Geographic (Latency-Optimized), and Geographic (Diversity-Optimized), evaluated across five simulated scenarios involving user loads from 250,000 to 1 million and network sizes from 10,000 to 50,000 relays. A total of 37,500 circuits were successfully generated with 100\% build success, enabling an in-depth evaluation of throughput, latency, efficiency, and scalability.

\begin{figure}[H]
\centering
\includegraphics[width=0.35\textwidth]{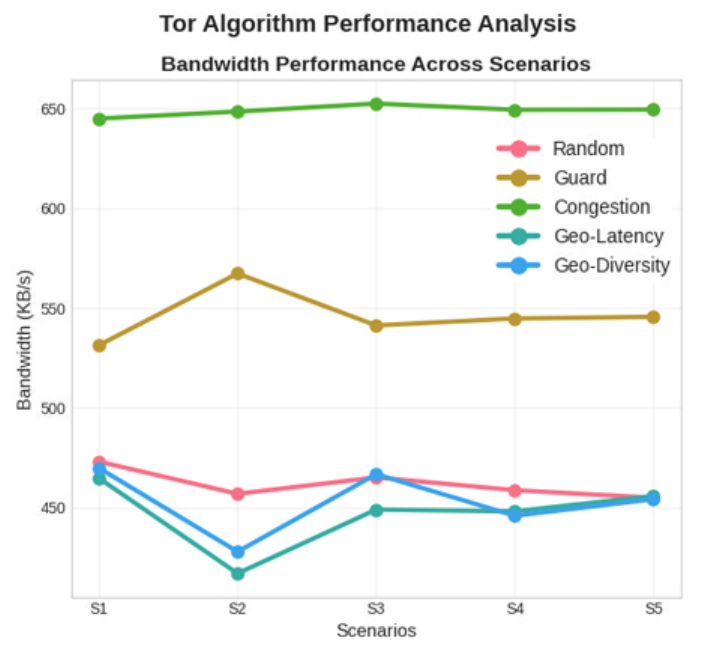}
\hfill
\caption{Bandwidth performance across scenarios.}
\label{fig:1i}
\end{figure}

\begin{figure}[H]
\centering
\includegraphics[width=0.35\textwidth]{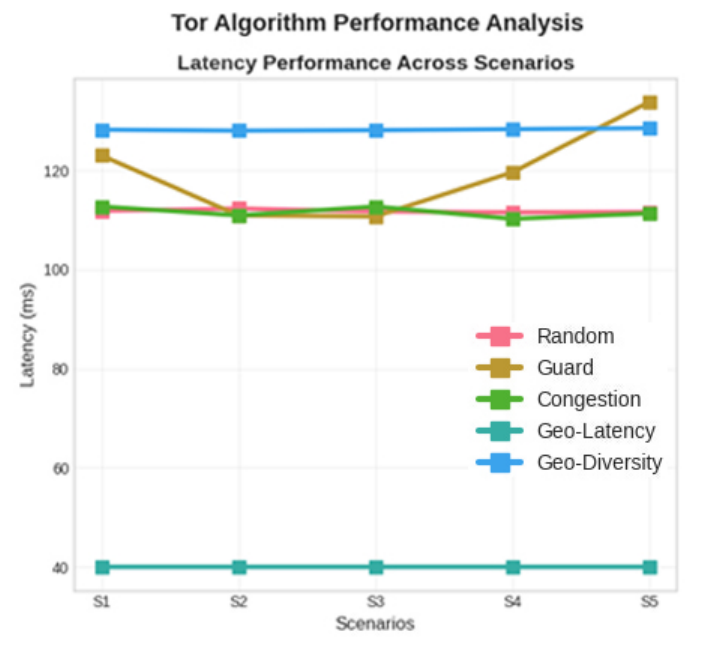}
\hfill
\caption{Latency performance across scenarios.}
\label{fig:1ii}
\end{figure}

\begin{figure}[H]
\centering
\includegraphics[width=0.35\textwidth]{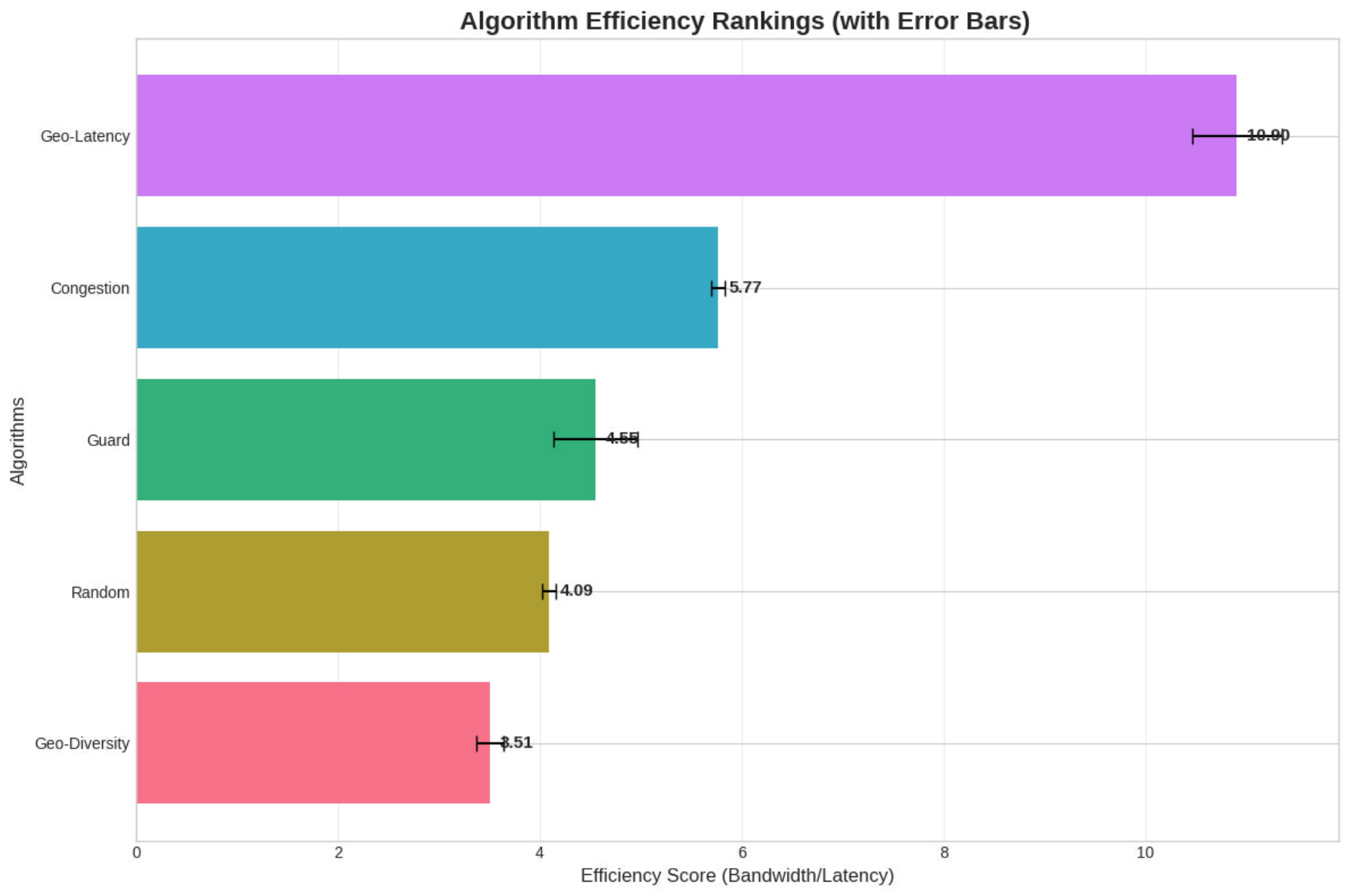}
\hfill
\caption{Algorithms efficiency ranking by bandwidth/latency.}
\label{fig:2}
\end{figure}

\begin{figure}[H]
\centering
\includegraphics[width=0.35\textwidth]{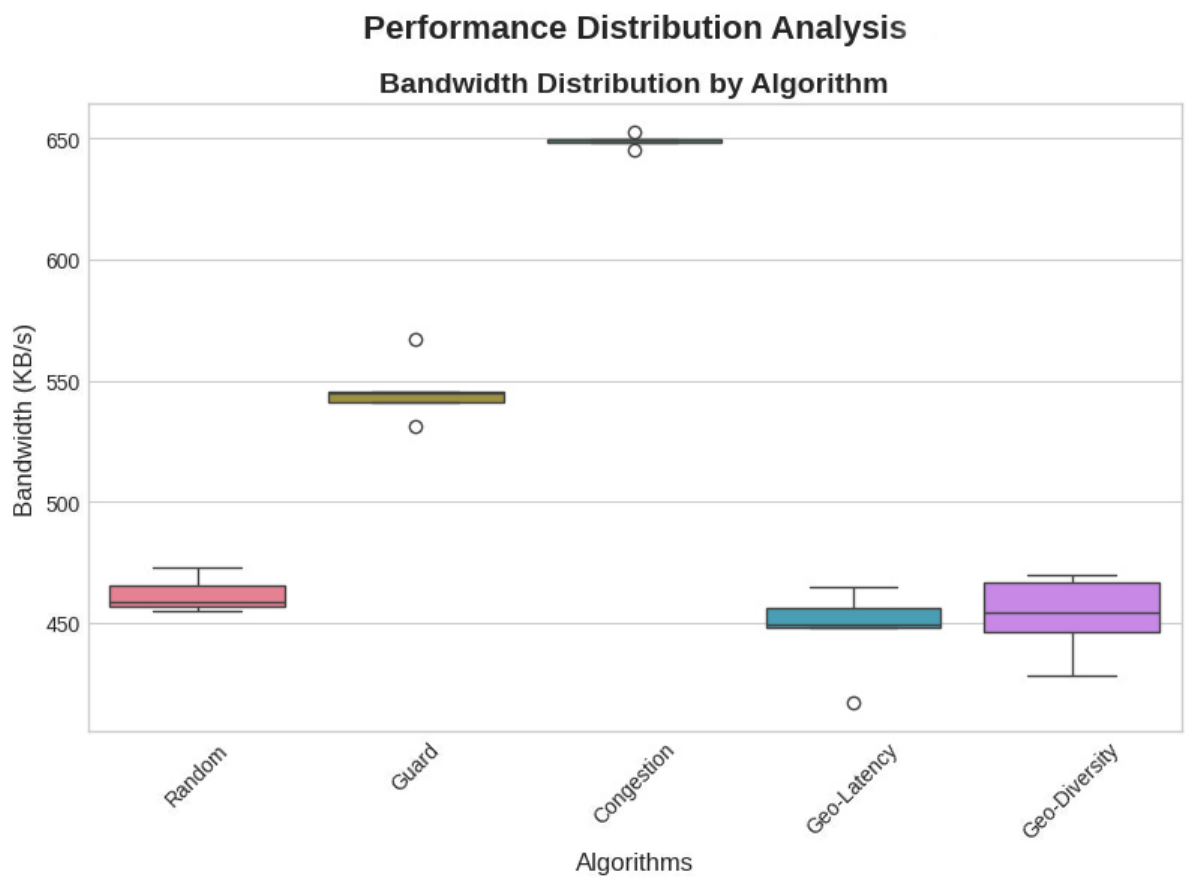}
\hfill
\caption{Bandwidth distribution by the various algorithms.}
\label{fig:4i}
\end{figure}

\begin{figure}[H]
\centering
\includegraphics[width=0.35\textwidth]{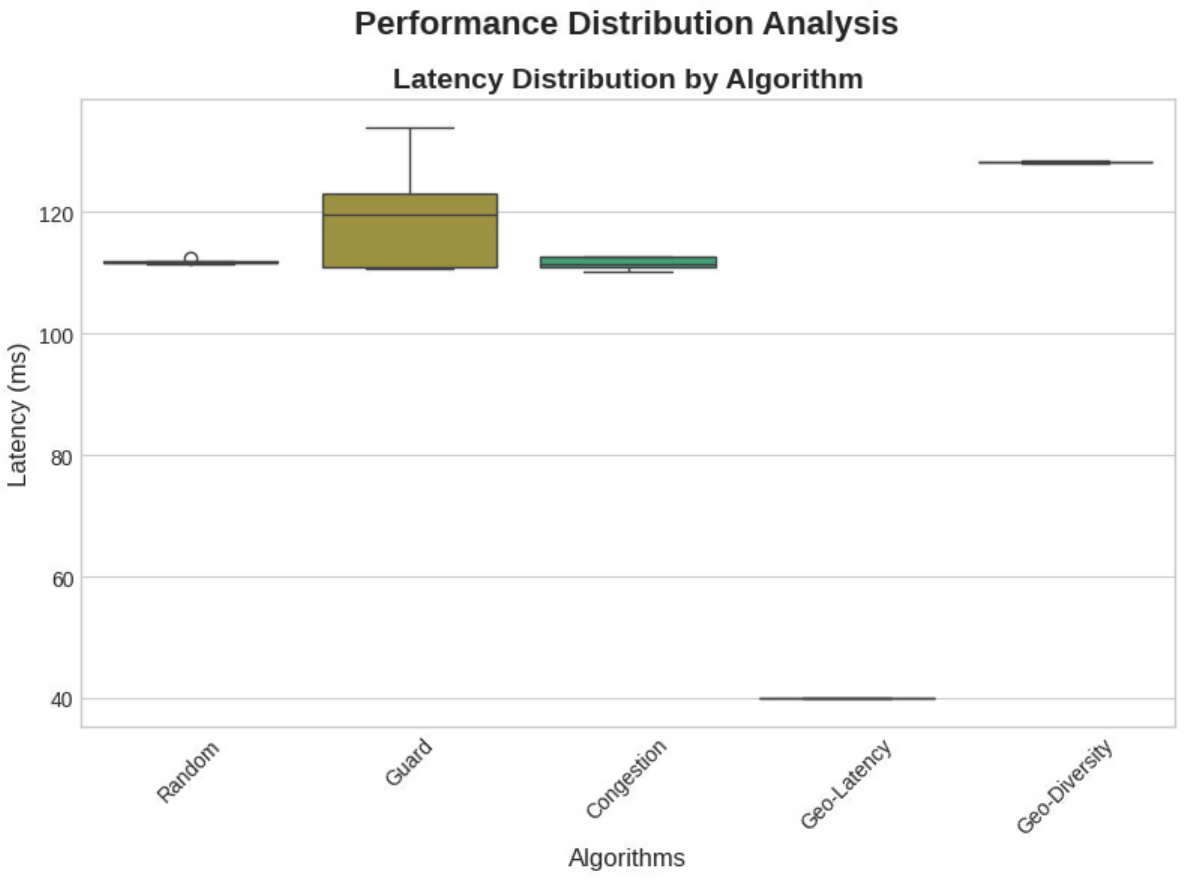}
\hfill
\caption{Latency distribution by various algorithms.}
\label{fig:4ii}
\end{figure}

Figures ~\ref{fig:1i} to ~\ref{fig:4ii} shows the bandwidth and latency performance alongside the resulting efficiency scores. Congestion-Aware achieved the highest throughput, with values ranging from 644.9 to 652.4 KB/s, representing a 35-42\% gain over the random baseline ( see Figures ~\ref{fig:1i} and ~\ref{fig:1ii}). In contrast, the Geographic Latency-Optimized (Geo-Latency) strategy maintained an exceptional constant latency of 40.0 ms across all scenarios (as shown in Figure~\ref{fig:1ii}), marking a 64-70\% reduction over other algorithms, averaging between 110 and 130 ms. These advantages translated into the highest efficiency score of 10.90, followed by Congestion-Aware (5.77), Guard (4.55), Random (4.09), and Geographic Diversity-Optimized (Geo-Diversity) as (3.51), shown in Figure~\ref {fig:2}. The boxplots (Figures ~\ref{fig:4i} and ~\ref{fig:4ii} ) show that Congestion-Aware exhibited minimal variation in throughput, and Geographic Latency-Optimized (Geo-Latency) retained absolute consistency in latency. In contrast, Guard and Geographic Diversity-Optimized (Geo-Diversity) showed wider performance spreads.

\begin{figure}[H]
\centering
\includegraphics[width=0.35\textwidth]{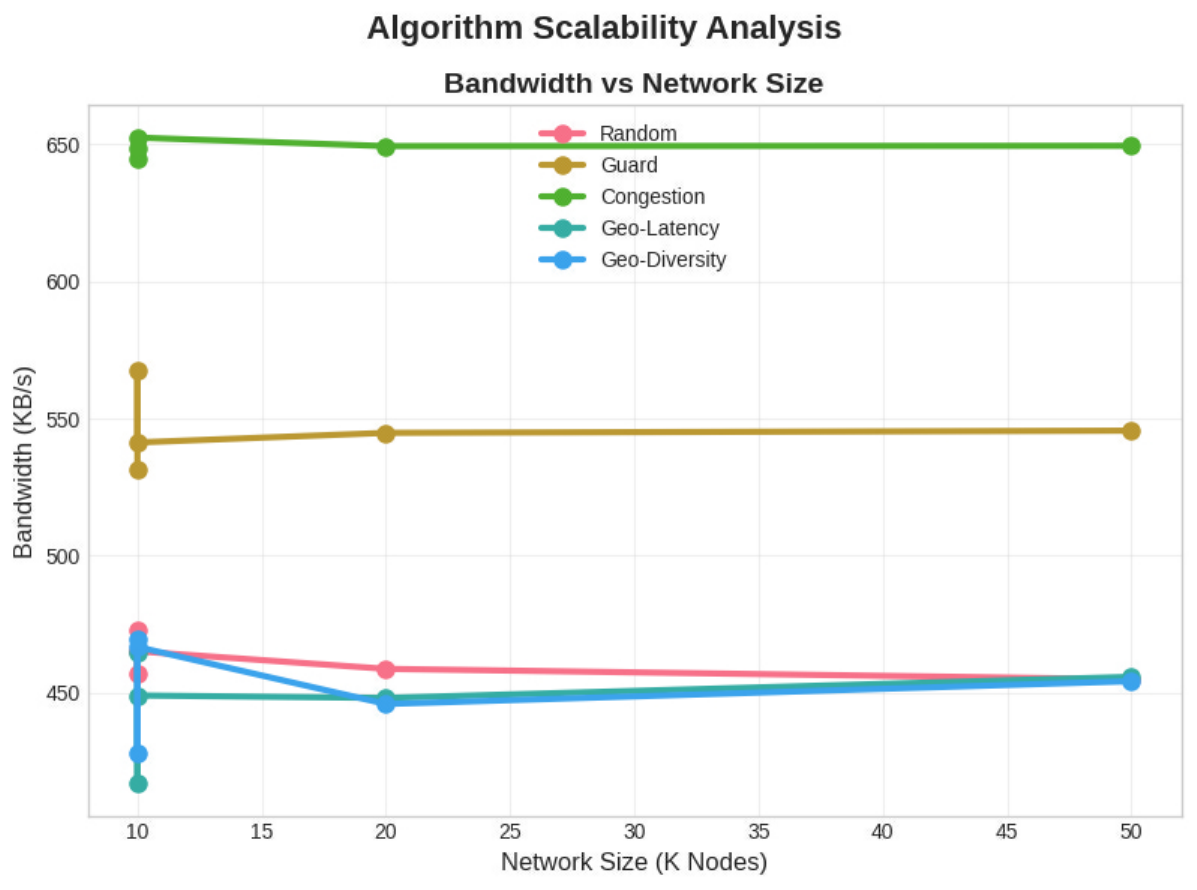}
\hfill
\caption{Algorithm scale on Bandwidth vrs Network Size.}
\label{fig:5i}
\end{figure}

\begin{figure}[H]
\centering
\includegraphics[width=0.35\textwidth]{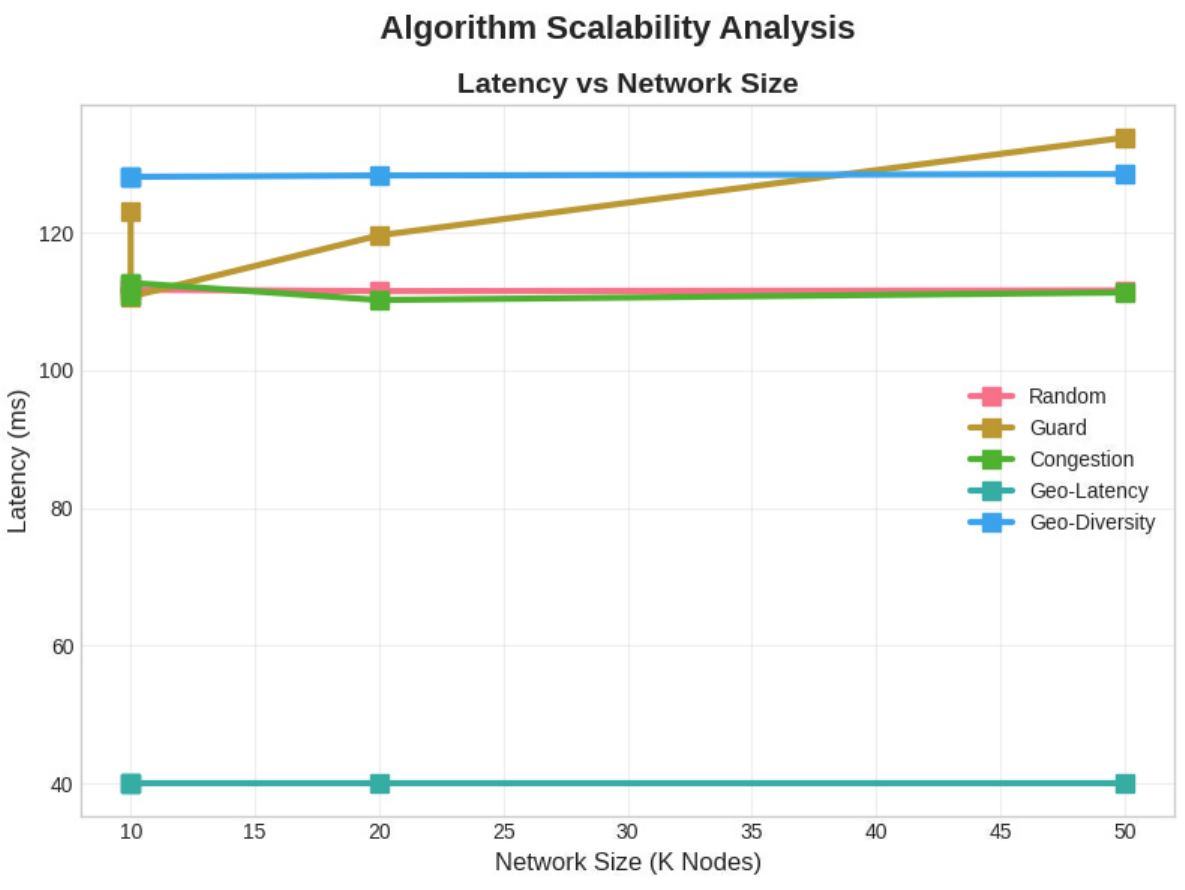}
\hfill
\caption{Algorithm scale on Latency vrs Network Size.}
\label{fig:5ii}
\end{figure}

\begin{figure}[H]
\centering
\includegraphics[width=0.35\textwidth]{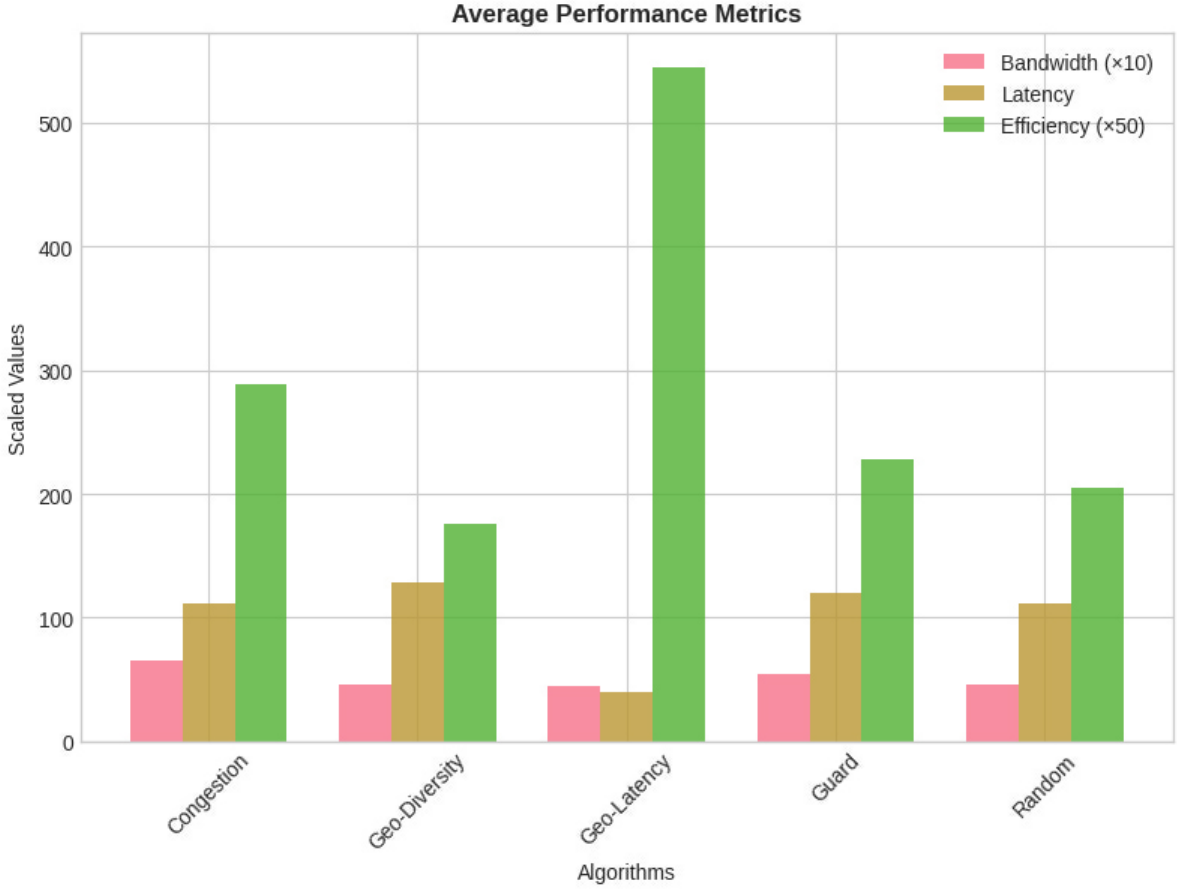}
\hfill
\caption{Algorithms average performance metrics.}
\label{fig:6i}
\end{figure}

\begin{figure}[H]
\centering
\includegraphics[width=0.35\textwidth]{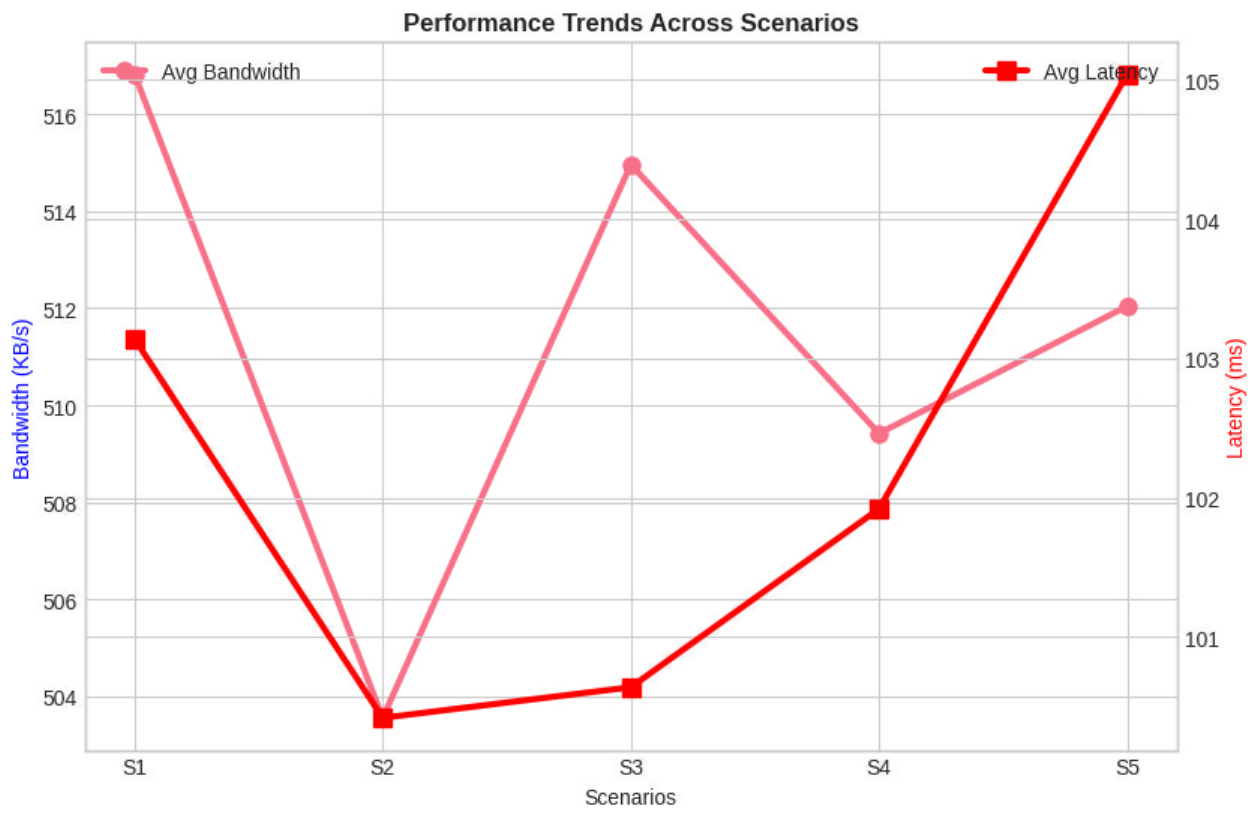}
\hfill
\caption{Performance trends across different scenarios.}
\label{fig:6ii}
\end{figure}

\begin{figure}[H]  % Use capital H from the float package
\centering
\includegraphics[width=0.35\textwidth]{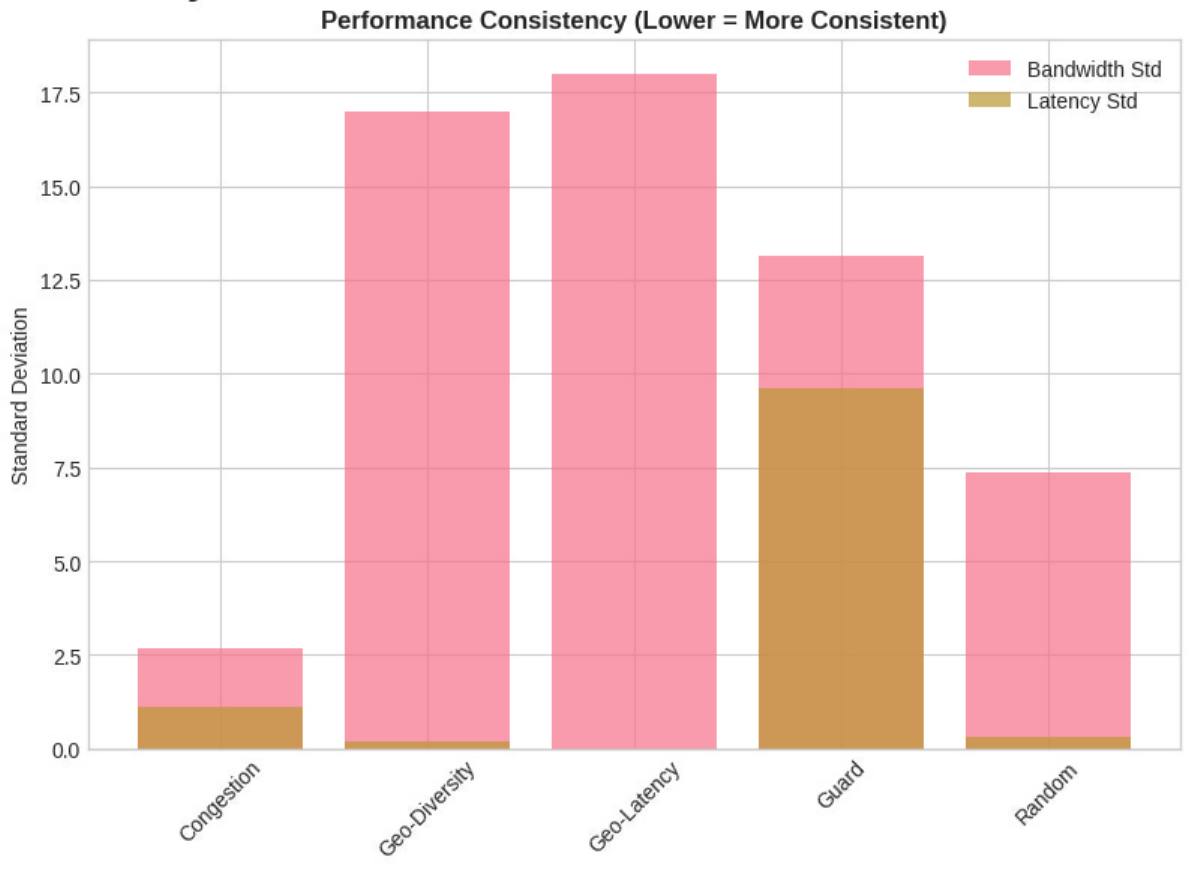}
\caption{Consistency of performance by standard deviation.}
\label{fig:6iii}
\end{figure}

Scalability analysis in Figures ~\ref{fig:5i} ,  ~\ref{fig:5ii}, ~\ref{fig:6i}, ~\ref{fig:6ii}, and ~\ref{fig:6iii} demonstrates how each algorithm performs as network size increases. While throughput remained stable for Congestion-Aware and Random (see ~\ref{fig:5i}), the Guard strategy experienced a steady rise in latency (123.0 to 133.8 ms) under scaling, indicating saturation of persistent guard nodes (see ~\ref{fig:5ii}). Geographic (Latency-Optimized), by contrast, retained its latency advantage at all scales. These results are consolidated in the summarization average metrics, standard deviation (consistency), scenario trends, and the top-performing algorithms (see ~\ref{fig:6i}, ~\ref{fig:6ii}, ~\ref{fig:6iii}). Random and Congestion-Aware exhibited the lowest performance variability, whereas the geographic methods showed higher variance.

\begin{figure}[H]
\centering
\includegraphics[width=0.35\textwidth]{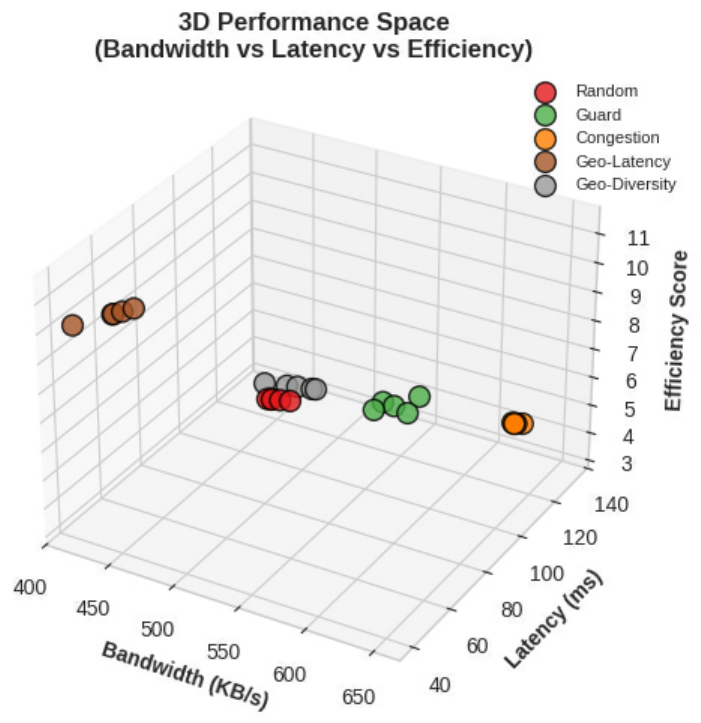}
\caption{3D performance space showing bandwidth, latency, and efficiency. Geo-Latency dominates the low-latency, high-efficiency zone.}
\label{fig:7i}
\end{figure}

\begin{figure}[H]
\centering
\includegraphics[width=0.35\textwidth]{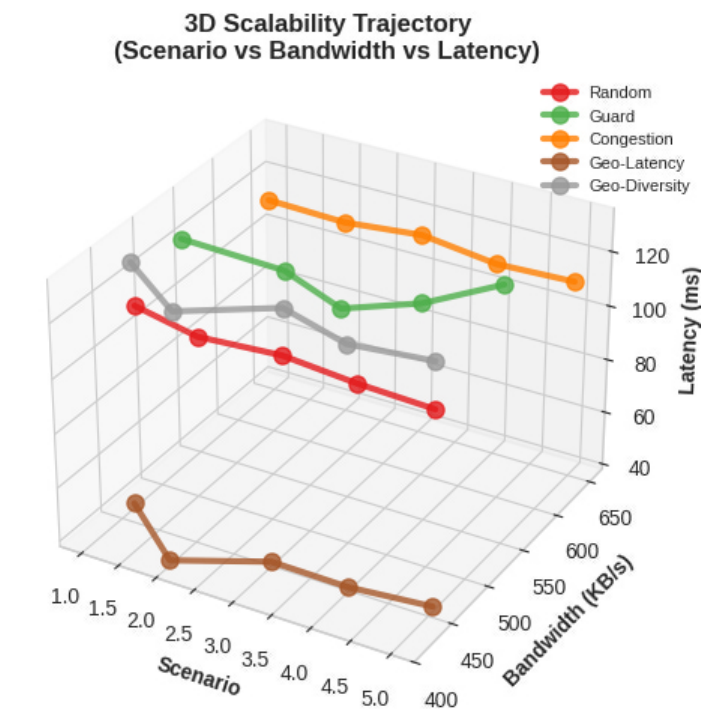}
\caption{Scalability trends across scenarios. Geo-Latency remains stable; others vary under increasing network size.}
\label{fig:7ii}
\end{figure}

\begin{figure}[H]
\centering
\includegraphics[width=0.35\textwidth]{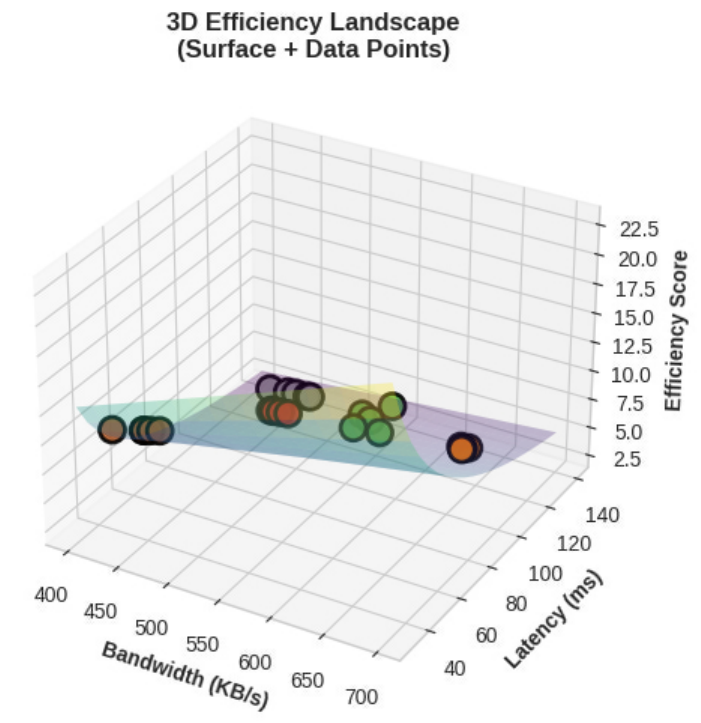}
\caption{Efficiency surface overlay. Geo-Latency aligns with the ridge; others drift under high-latency or low-bandwidth.}
\label{fig:7iii}
\end{figure}

\begin{figure}[H]
\centering
\includegraphics[width=0.35\textwidth]{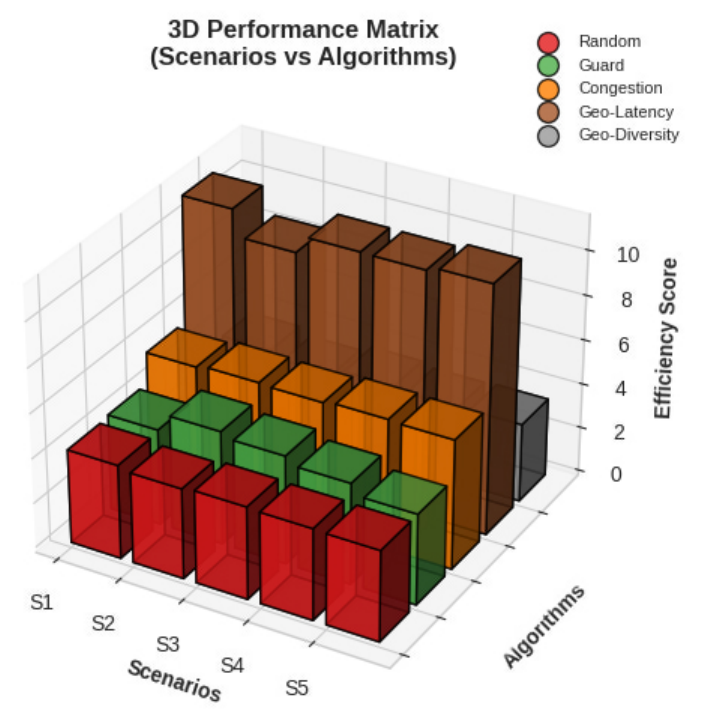}
\caption{Per-scenario efficiency matrix. Geo-Latency leads across all scales; Geo-Diversity ranks lowest.}
\label{fig:7iiii}
\end{figure}

\begin{figure}[H]
\centering
\includegraphics[width=0.35\textwidth]{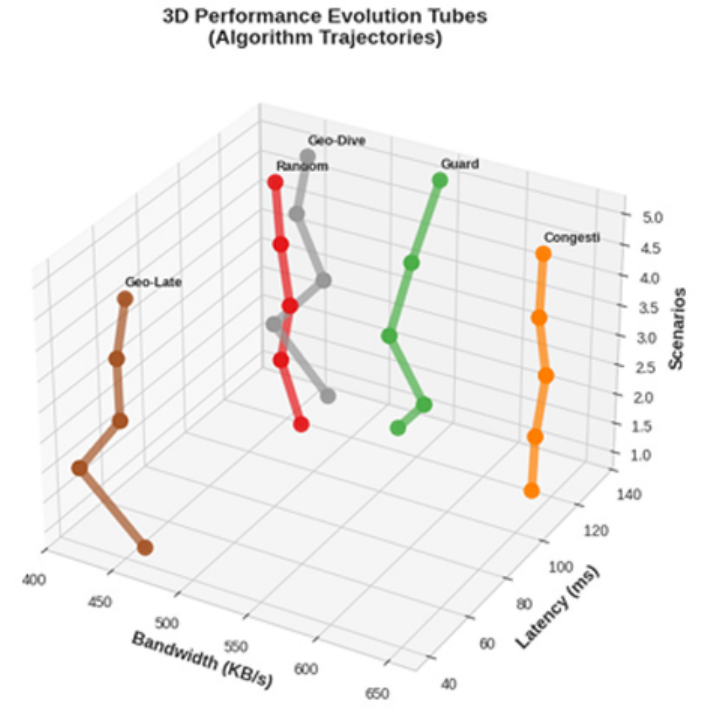}
\caption{3D performance tubes across scenarios. Geo-Latency improves steadily; others plateau or fluctuate.}
\label{fig:7iiiii}
\end{figure}

\begin{figure}[htbp]
\centering
\includegraphics[width=0.35\textwidth]{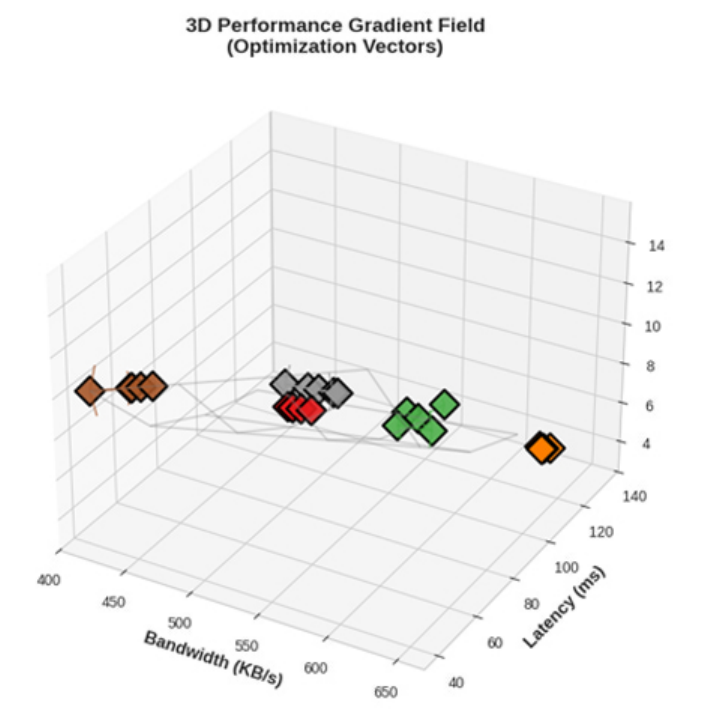}
\caption{Optimization vectors in 3D space. Geo-Latency and Congestion-Aware follow strong performance gradients.}
\label{fig:7iiiiiii}
\end{figure}

\begin{figure}[htbp]
\centering
\includegraphics[width=0.35\textwidth]{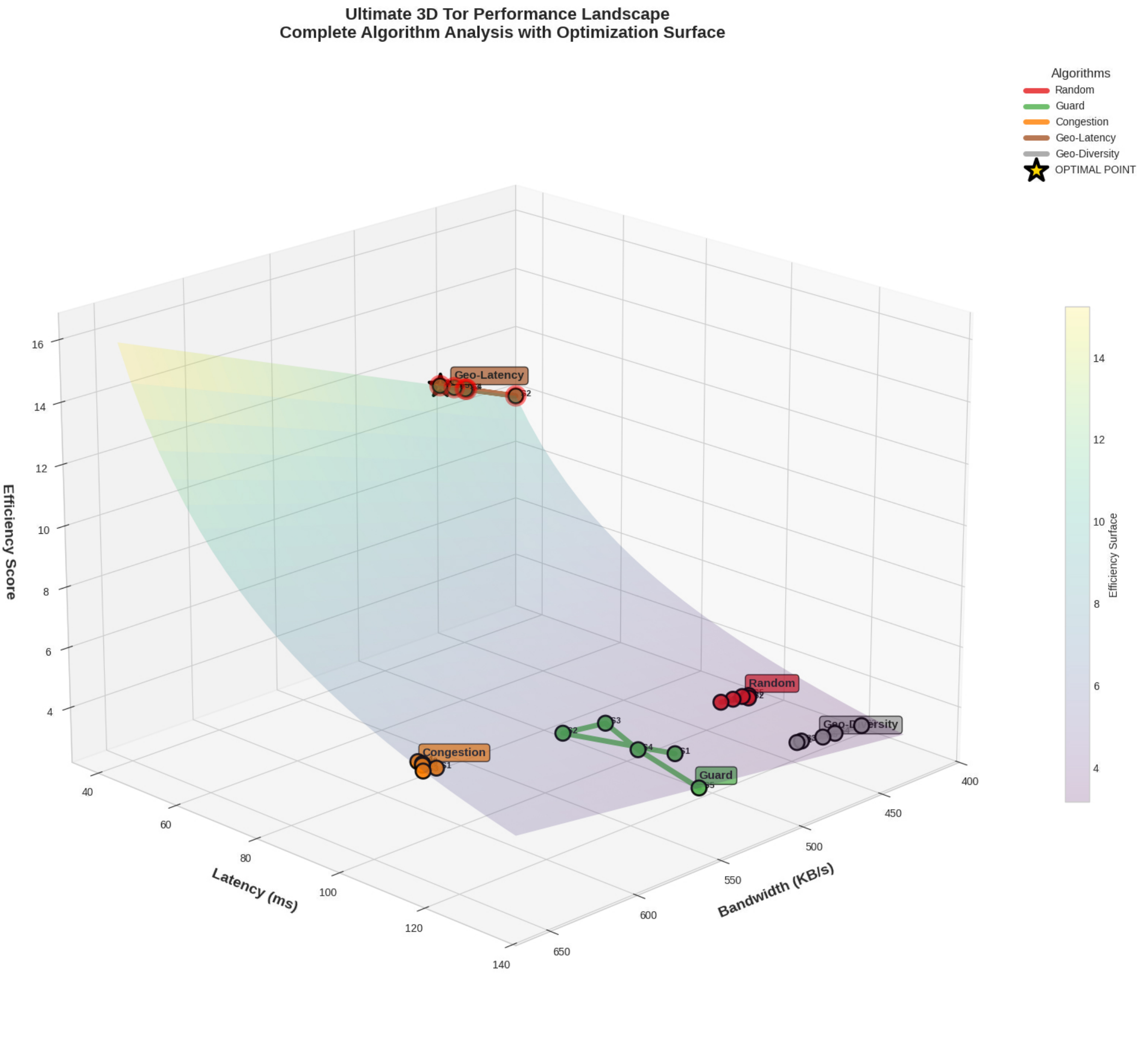}
\caption{Global optimization surface. Geo-Latency aligns with peak efficiency; others diverge from optimal path.}
\label{fig:7iiiiiiii}
\end{figure}

To provide a clearer, more granular understanding of how each algorithm behaves across key performance dimensions, we decomposed the composite three dimensional visualization into distinct 3D figures (Figures~\ref{fig:7i} – \ref{fig:7iiiiiiii}). Each captures a specific analytical perspective, offering a comprehensive view of bandwidth, latency, efficiency, and scalability trade-offs.

Figure~\ref{fig:7i} presents the core performance space where bandwidth, latency, and efficiency scores are jointly visualized. The Geographic Latency-Optimized (Geo-Latency) strategy occupies the high-efficiency, low-latency region, while Congestion-Aware aligns with the high-bandwidth zone. Guard and Random strategies cluster toward the center, suggesting moderate and balanced performance. Geo-Diversity appears more dispersed, highlighting its inconsistency. In Figure~\ref{fig:7ii}, we observe the scalability trajectory across five network scenarios. Geo-Latency maintains low latency and stable performance as network size increases, underscoring its robustness. Congestion-Aware offers consistently high bandwidth but at the cost of rising latency. Guard, Random, and Geo-Diversity exhibit more variable paths, with Geo-Diversity being the least stable.

Figure~\ref{fig:7iii} overlays an efficiency surface on the bandwidth-latency space. Geo-Latency consistently traces the peak of this surface, indicating strong optimization. Congestion-Aware remains near the upper plane but diverges slightly in high-latency zones. Random and Guard are positioned midway, while Geo-Diversity trails behind on the efficiency scale. Figure~\ref{fig:7iiii} visualizes the per-scenario efficiency matrix across all five strategies. Geo-Latency consistently achieves the highest scores in every scenario. Congestion-Aware follows with competitive values. Guard and Random remain in the middle range, and Geo-Diversity once again ranks lowest, reflecting its performance volatility.

In Figure~\ref{fig:7iiiii}, performance evolution tubes trace how each algorithm changes over time and scale. Geo-Latency shows a steady, upward trend in efficiency, while Guard and Congestion-Aware move vertically then plateau, suggesting performance saturation. Random and Geo-Diversity display irregular, nonlinear trajectories. 

Figure~\ref{fig:7iiiiiii} depicts optimization vectors, revealing each algorithm's directional movement through performance space. Geo-Latency and Congestion-Aware follow smooth, ascending gradients, suggesting convergence toward optimal trade-offs. Guard and Random have shorter, more static vectors, while Geo-Diversity’s scattered directionality indicates inefficiency and unpredictability.

Finally, Figure~\ref{fig:7iiiiiiii} offers a comprehensive landscape of all algorithms projected onto an optimization surface. Geo-Latency is consistently aligned with the peak efficiency zone. Congestion-Aware’s path is slightly below but steady. Other strategies diverge more significantly, affirming the superiority of Geo-Latency for latency-critical scenarios and Congestion-Aware for throughput-oriented use cases.

Together, these individual visualizations reveal that while no algorithm universally dominates, each presents a unique optimization profile. Geo-Latency is the most consistent and efficient across conditions, Congestion-Aware provides exceptional bandwidth, and Guard maintains balance with slight latency costs. Random and Geo-Diversity are less predictable and perform better in general-purpose rather than specialized use cases.

\section{Conclusion}\label{sec5}

This study evaluated five Tor path selection strategies across varied network conditions to understand their performance trade-offs. Using a high-fidelity simulation framework, we analyzed 37,500 circuits under realistic scaling scenarios to assess throughput, latency, efficiency, and scalability. The findings reveal that while no algorithm performs best across all metrics, distinct strengths make each suitable for different operational needs.

Geographic (Latency-Optimized), also known as Geo-Latency, consistently delivered the lowest latency of 40.0 ms in every scenario, and achieved the highest efficiency score (10.90), making it the best choice for latency-sensitive applications. Congestion-Aware, by contrast, excelled in throughput, outperforming the Random baseline by 35 to 42\%, and maintaining stable efficiency, making it ideal for bandwidth-intensive use cases. Guard-based selection provided moderate performance with reliable security properties but increased latency as the network scaled, indicating limited scalability. Random and Geo-Diversity approaches offered balanced yet suboptimal outcomes.

Scalability results confirmed that Geographic (Latency-Optimized) maintained its performance regardless of network size, while other strategies showed minor degradation. The 3D analysis reinforced these conclusions, highlighting each algorithm’s trajectory in the bandwidth-latency-efficiency space and confirming their distinct optimization paths.

These results suggest that tailoring path selection to the demands of specific applications can enhance Tor's usability without compromising anonymity. Latency-aware routing offers significant gains for interactive or time-sensitive services. For large data transfers, congestion-sensitive paths improve throughput. Default strategies remain viable for general use but are no longer the most efficient.

\section{Limitations and Future Work}\label{sec6}

While this study provides a detailed comparative analysis of Tor path selection strategies under realistic network conditions, several limitations should be acknowledged. First, the simulation framework, though high-fidelity, operates in a controlled environment and may not fully capture the unpredictability of live Tor traffic, relay churn, and user behavior. Second, geographic modeling was limited to four broad regions, which may oversimplify real-world routing complexities and ignore finer variations in latency or jurisdictional boundaries. Third, the study did not explicitly evaluate resilience against adversarial threats \cite{kornyo2024enhancing} such as relay compromise, traffic correlation, or network partitioning.

Future work should build on these findings in several directions. Adaptive hybrid algorithms that switch between selection strategies based on real-time network states and application needs could offer more flexible performance optimization. For example, latency-sensitive users could benefit from Geographic (Latency-Optimized) paths, while bandwidth-heavy tasks could automatically switch to Congestion-Aware routes.

Expanding the geographic model to include finer-grained locations, autonomous system diversity, and legal jurisdiction separation would enhance both the anonymity set and performance granularity. Such improvements could also support routing policies that respect political or legal boundaries, adding an extra layer of safety for users in hostile environments.

Security-focused evaluation under adversarial conditions is also essential. Future research should simulate targeted attacks, relay misbehavior, and correlation attempts to assess the robustness of each strategy. Understanding how algorithms behave under stress will ensure anonymity guarantees are not undermined.

Finally, deploying these strategies in a live Tor environment is necessary for practical adoption. Real world testing with active relays and user traffic would reveal operational constraints and opportunities not visible in simulation. Machine learning techniques, informed by historical path performance and user profiles, may also offer promising avenues for intelligent, context-aware routing while preserving user anonymity through privacy-preserving learning methods.

Addressing these areas will further strengthen Tor's ability to deliver security and performance across various applications and network

%
% ---- Bibliography ----
%
\bibliographystyle{IEEEtran}
\bibliography{refs}

% ---- OR ------- Uncomment this section to use bibtex
% \bibliographystyle{spmpsci} % We choose the "plain" reference style
% \bibliography{refs} % Entries are in the refs.bib file
\end{document}